\documentclass[11pt]{article}
\usepackage[utf8]{inputenc}

\usepackage{graphicx}
\usepackage{appendix}

\usepackage[margin=1.25in]{geometry}
\usepackage{url}

\newcommand{\ph}[1]{\medbreak \noindent {\bf #1}}

\newcounter{srctr}
\def\srule#1{\refstepcounter{srctr}%
  {\medbreak\noindent {\bf Lesson \thesrctr. {#1} }}%
}

\title{Ten Lessons for Data Sharing \protect\\ With a
Data Commons}

\author{Robert L. Grossman\protect\\
Center for Translational Data Science\protect\\ 
University of Chicago}
\date{June 2022}

\begin{document}

\maketitle

\section*{Abstract}
A data commons is a cloud-based data platform with a governance structure that allows a community to manage, analyze and share its data.  Data commons provide a research community with the ability to manage and analyze large datasets using the elastic scalability provided by cloud computing and to share data securely and compliantly, and, in this way, accelerate the pace of research. Over the past decade, a number of data commons have been developed and we discuss some of the lessons learned from this effort.

\section{What is a Data Commons?}
{\bf Cloud computing has made it much easier to store and analyze large datasets, but it is still a challenge to share data effectively with a research community in a way that accelerates research.}

\medbreak

Since the end of the 16th century, experimental science has driven by conducting experiments, collecting data, and analyzing it. With the development of low cost sensors, high throughput instruments, and inexpensive storage to hold the data they produce, a new paradigm for scientific discovery has emerged.  The {\em sharing of  data}, and its reanalysis in different contexts, with different aims, and in conjunction with other datasets to create new hypotheses and new scientific discoveries is part of what is sometimes called the fourth paradigm of science \cite{hey2009fourth}.\footnote{The other three paradigms are: experimental science, theoretical science and simulation science.}   In this article, we discuss the role of software platforms called data commons in supporting the fourth paradigm of science and some lessons learned from ten years of experience developing and operating data commons.

As we will describe below, a data commons is a shared resource to support a scientific community.  
Some of the challenges with shared resources were identified in 1968 when Joseph Hardin published an article in Science called the {\em The Tragedy of the Commons} \cite{hardin1968tragedy} that focused attention on problems arising when a shared finite resource is used by a community. The governance structure is critical.  About forty years later in 2009, Elenor Ostrom received the Nobel prize in Economic Sciences \cite{janssen2012elinor} for her work about the governance of the commons \cite{ostrom1990governing}.

Following Ostrom, we define a {\em commons} as a natural, cultural or digital resource accessible to all members of a community, or more broadly of a society. Importantly, a commons is held through a partnership, a not-for-profit, or other entity, for the benefit of a community, but not owned privately for commercial gain \cite{ostrom1990governing}.

For the purposes here, we view a {\em data commons} as a software platform that co-locates: 1) data, 2) cloud-based computing infrastructure, and 3) software applications, tools and services to create a governed resource for managing, analyzing, and sharing data with a community \cite{grossman2019data}. 
{\bf Briefly, a data commons is a cloud-based software platform with a governance structure that allows a community to manage, analyze and share its data.}

Examples of data commons and similar platforms include: 
the NCI Genomic Data Commons (GDC) \cite{heath2021GDC}, 
the NHLBI BioData Catalyst data platform \cite{BDC},  
the NHGRI Genomic Data Science Analysis, Visualization and Informatics Lab-space (AnVIL) \cite{schatz2022inverting}, the NIH Common Fund Data Ecosystem \cite{white2021making}, the All of Us Research Hub \cite{all2019all},
the BloodPAC Data Commons \cite{grossman2021bloodpac},
the NIBIB Medical Imaging and Data Resource Center (MIDRC) \cite{giger2021medical}, the Veterans Precision Oncology Data Commons \cite{do2019veterans}, and the NIH Kids First Data Resource \cite{KF}. 

\section{Why Build Data Commons?}

There are several main reasons research projects build data commons.

\begin{enumerate}

\item {\bf The functionality is compelling.}  Modern cloud computing provides elastic, on-demand, pay-as-you-go computing that can be used to provide compelling functionality in a data commons and accelerate research over the data in the commons. The most important reason for building a data commons is that the functionality provided by a data commons is compelling.  

\item {\bf To speed the pace of research discoveries.} By having the commons curate and processes the data once for a particular research community, it enables  individual researchers and research groups to  proceed more quickly to analyzing data to investigate particular hypotheses.  This increases the pace of research discoveries. 

\item {\bf To create network effects.}. Another reason to build data commons is be part of a larger data ecosystem or data mesh containing multiple commons, computing platforms, and knowledgebases in order to take advantage of network effects as more commons are built and more users access data from the resources in the mesh. As an example, a commons in a data mesh can participate in federated machine learning with other commons and data resources via its APIs, while adhering to its governance, security and privacy policies. 

\item {\bf To host data that is too large to be managed easily by research groups.}  As the size of the data grows, it becomes more and more difficult for each research group to develop and operate their own computing infrastructure.  A data commons is often used to manage the large experimental data and process it to produce derived datasets that are easier to analyze by individual scientists and research groups.

\item {\bf To reduce cost.} Funds for science are always limited and tough choices must be made.  By investing in a centralized commons, the cost for managing, curating and processing data to produce data products for a research community and can be done once centrally to reduce the overall costs of a research program or initiative.   In practice, with very large datasets, the cost of funding each research group to build their own data platform can be so high that the only practical way to distribute the data to the research community is through a centralized data commons.  

\item {\bf To protect sensitive data.} Sometimes data commons are built because research data is so sensitive that it must be protected within enclave for it to be safely shared.  Examples include the All of Us Research Hub \cite{all2019all} and the Veterans Precision Oncology Data Commons \cite{do2019veterans}.

\end{enumerate}

As an example, data commons, such as the NCI Genomic Data Commons \cite{heath2021GDC, zhang2021uniform}, have made petabytes of curated, harmonized data availability to a research community and ``democratized access to cancer genomics data,'' which until the launch of the GDC in 2016 was available only to the largest research organizations that had the resources and expertise to analyze petabyte scale data.  With the GDC and its open APIs \cite{wilson2017developing}, researchers could access the GDC's data products in cloud computing environments or on their own local resources for further analysis or integrative studies.

The emergence and adoption of cloud computing over the past decade \cite{stein2010case} has made developing data commons and data meshes much easier.  Today, cloud computing is the key technology that has enabled the current generation of data commons.

\ph{A data gap.} Although the large and increasing amount of data being generated in biology, medicine and healthcare is well documented \cite{bell2009beyond,hey2009fourth,stein2010case}, the amount of well-curated, harmonizend data is often not noted.   This creates a  ``data gap'' that impedes research.  Data commons are designed in part to close this gap and allow research communities to create curated, harmonized data sets to accelerate their research.

\section{The Success of the NCI Genomic Data Commons}

One of the more popular data commons is the NCI Genomic Data Commons \cite{heath2021GDC, zhang2021uniform}.  As reported in \cite{heath2021GDC}, the GDC contains over 2.9 PB of curated, harmonized cancer genomics data from over 60 projects (as of February 2021).  Each month over 50,000 unique researchers use the system and over 1.5 PB of data are accessed \cite{heath2021GDC}.  As a rough measure of its popularity, the membership of the American Association of Cancer Research is about 50,000.

The data from the different projects is curated with respect to a single data model and each month on average over 25,000 bioinformatics pipelines are run over the data to create a harmonized set of data products that are analyzed with a common set of bioinformatics pipelines \cite{zhang2021uniform}. In contrast, it is typical to bring together cancer genomics data from different projects that are analyzed by different groups using different pipelines, which can make subsequent integrative analysis much more challenging and problematic.  The GDC has a user interface that enables interactive graphical exploration of the data, with the ability to download publication quality graphics.  

Perhaps the most important reason for its popularity is that the GDC makes it easy for researchers to access its data and make new research discoveries with much less effort than if they were to analyze the raw data themselves, as it usually required with a traditional data repository. 

The GDC lists over 100 high impact publications that have been written using the data products it makes available to the cancer genomics research community.

\section{Ten Simple Rules}

\srule{Build a commons for a specific community with a specific set of research challenges.} Although there are a few data repositories that serve the general scientific community that have proved successful, in general data commons that target a specific user community have proven to be the most successful.  The first lesson is to build a data commons for a specific research community that is struggling to answer specific research challenges with data. As a consequence, a data commons is a partnership between the data scientists developing and supporting the commons and the disciplinary scientists with the research challenges.

\srule{Successful commons curate and harmonize the data.} Successful commons curate and harmonize the data and produce data products of broad interest to the community.  It's time consuming, expensive, and labor intensive to curate and harmonize data, by much of the value of data commons is centralizing this work so that it can be done once instead of many times by each group that needs the data.  These days, it is very easy to think of a data commons as a platform containing data, not spend the time curating or harmonizing it, and then be surprised that the data in the commons is not used more widely used and its impact is not as high as expected.

\srule{It's ultimately about the data and its value to generate new research discoveries.}  Despite the importance of a study, few scientists will try to replicate previously published studies.  Instead,  data is usually accessed if it can lead to a new high impact paper.  For this reason, data commons play two different but related roles.  First, they preserve data for reproducible science.  This is a small fraction of the data access, but plays a critical role in reproducible science.  Second, data commons make data available for new high value science.

\srule{Reduce barriers to access to increase usage.}  A useful rule of thumb is that every barrier to data access cuts down access by a factor of 10.  Common barriers that reduce use of a commons include: registration vs no-registration; open access vs controlled access; click through agreements vs signing of data usage agreements and approval by data access committees; license restrictions on the use of the data vs no license restrictions.
    
\srule{Data curation and developing interactive user interfaces is expensive.} The largest costs of developing and operating a data commons are: i) the costs of data curation and data harmonization; and, ii) the costs of developing easy to use, interactive front ends for exploring and analyzing the data.

\srule{Support an ecosystem of applications, not just a single system.} The most successful commons have open APIs, enabling  the community to build  utilities, libraries, and tools make the data usable and accessible. Even better is when the sponsor of a commons provides direct support for community efforts to develop applications over the open APIs of a commons.  In contrast, some platforms are developed with the goal of keeping all the data in the platform and forcing the community to work entirely within the platform.  In between, are data platforms that interoperate with other data platforms that they trust, so that data can be shared, integrated and analyzed within the community of trusted data platforms.

\srule{Security and compliance for data commons are expensive.} The policies, procedures and controls required for security, compliance, and regulatory support are expensive and time consuming to develop and to operate and, in almost all cases, are underfunded.  

\srule{It's not easy to predict what archived data will lead to great science.} Some datasets in a common tend to be very frequently downloaded, with others much less frequently downloaded, and with the distribution following a Zipfian or other power law distribution \cite{clauset2009power}.  This might lead one to advocate saving operational costs by only hosting the more popular datasets.  On the other hand, over the long term, interesting, and sometimes, quite interesting, new science may result from further use and analysis of the less popular datasets.

\srule{Over time, the value of data commons will grow if it is part of a data mesh.} Data commons, at least as we have defined them here, are designed to support a particular research community.  In contrast, a data mesh (also known as a data ecosystem) is a hybrid architecture, consisting of some common services that enable a collection of independently managed and governed data platforms to interoperate.  By using some common services for authentication, authorization and data access, a data commons can also enable other research communities to access their data for integrative and cross disciplinary analysis \cite{grossman2018progress}.

\srule{Resist the temptation to build a cloud-based walled garden.}  It is tempting when building a data platform or data commons to take the attitude that everyone should contribute data to your platform, but there is not a good reason, and only risk, if you enable your data to leave your system. This is especially the case when data is sensitive.   It is equally tempting to take the view that all analysis should be done within your platform.  Both urges should be resisted.  As an alternative, if data access needs to be restricted, consider developing trust relationships with other data platforms and sharing data and interoperating with them.

\section{Summary and Conclusion}

Biology, medicine and healthcare are creating large datasets and cloud computing has provided the computing infrastructure to manage and analyze the resulting data \cite{stein2010case}.   A data commons is a cloud-based computing resource with a governance structure that allows a community to manage, analyze and share its data.

Data commons have emerged in part to address the data gap --- the gap between the large amount of data available and the small amount of data that can be easily used to formulate new hypotheses, to make new discoveries, and to build machine learning and AI models.  

This point is worth emphasizing.  {\bf Despite, the availability of raw generated data and large scale cloud computing infrastructure, biology, medicine and healthcare remain {\em data limited}, since the available data needs to be carefully curated and harmonized before it is useful. Commons support this important activity so that research questions can be more efficiently tackled by a research community.}

Looking towards the future, a core set of microservices are emerging \cite{rehm2021ga4gh} that enable data meshes (data ecosystems) consisting of multiple data commons, cloud-based computing platforms, knowledgebases, and other resources to interoperate.  The hope is that making data also available through interoperating with data meshes will further accelerate the pace of research.

\section*{Acknowledgements}
Research reported in this publication was supported by the NIH Common Fund under Award Number U2CHL138346, which is administered by the National Heart, Lung, And Blood Institute of the National Institutes of Health. The content is solely the responsibility of the author and does not necessarily represent the official views of the National Institutes of Health.

The author would like to thank Warren Kibbe for the helpful remarks and suggestions that he made about the paper.

\bibliographystyle{plain}

\clearpage

\appendix

\section{Glossary}
The definitions in this section are from \cite{grossman2019data} or \cite{grossman2018progress} or the references indicated.

\ph{Commons.} A commons is a natural, cultural or digital resource accessible to all members of a community, or more broadly of a society. Examples include a pasture for animals to graze in a village, a dog park for dogs in a city neighborhood, or natural materials such as air, water for society in general. These resources are held in common, through a partnership, a not-for-profit, or other entity, but not owned privately for commercial gain \cite{ostrom1990governing}.

\ph{Data clouds.} A data cloud is a cloud computing platform for managing, analyzing and sharing datasets.

\ph{Data commons.} A {\em data commons} as a software platform that co-locates: 1) data, 2) cloud-based computing infrastructure, and 3) software applications, tools and services to create a governed resource for managing, analyzing, and sharing data with a community \cite{grossman2019data}. Briefly, a data commons is a cloud-based software platform with a governance structure that allows a community to manage, analyze and share its data.  A data commons provides services so that the data is findable, accessible, interoperable and reusable (FAIR) \cite{wilkinson2016fair}.  With these services, a data commons is an interoperable resource. 

\ph{Data ecosystems.} Data ecosystems contain multiple data commons, data repositories, cloud computing resources, knowledgebases, and other applications that can interoperate using a common set of services (sometimes called framework services) for authentication, authorization, accessing, and analyzing data.  Data ecosystems are beginning to be called data meshes.

\ph{Data harmonization.}   Data harmonization as the process that brings together data from multiples sources and applies uniform and consistent processes, such as uniform quality control metrics to the accepted data; mapping the data to a common data model; processing the data with common bioinformatics pipelines; and post-posting the data using common quality control metrics.

\ph{Data lake.} A data lake is a system for storing data as objects, where the objects have an associated GUID and (object) metadata, but there is no data model for interpreting the data within the object.

\ph{Data meshes.} A data mesh is another name for data ecosystem.  See data ecosystem.

\ph{Data object.} In cloud computing, a data object consists of data, a key, and associated metadata.   The data can be retrieved using key and the metadata associated with a specific data object can be retrieved, but more general queries are not supported.  Amazon’s S3 storage system is a widely used storage system for data objects.  Compare to structured data.

\ph{Data portal.} A data portal is a website that provides interactive access to data in an underlying data management systems, such as a database.  Data commons, data lakes can also have data portals. 

\ph{FAIR data.} FAIR data are data which meet principles of findability, accessibility, interoperability, and reusability \cite{wilkinson2016fair}.

\ph{Framework services.}  A set of standards-based services with open APIs for authentication, authorization, creating and accessing FAIR data objects, and for working with bulk structured data in machine readable, self-contained format.  Framework services are used for build data commons and data ecosystems.  Framework services are example of the end-to-end design principle used in the core services underling the internet \cite{saltzer1984end}.  Framework services are also called mesh services.

\ph{Globally Unique Identifier (GUID).} A GUID is an essentially unique identifier that is generated by an algorithm so that no central authority is needed, but rather different programs running in different locations can generate GUID with a low probability that they will collide.  A common format for a GUID is the hexadecimal representation of a 128 bit binary number. 

\ph{Mesh Services.} Mesh services are another name for framework services.

\ph{Structured data.} Data is structured if it is organized into records and fields, with each record consisting of one or more data elements (data fields).   In biomedical data, data fields are often restricted to controlled vocabularies to make querying them easier.  Compare to data object.



\end{document}